\documentstyle[floats,tighten,prl,aps,epsfig,psfig]{revtex}
\begin{document}
\draft

\twocolumn[\hsize\textwidth\columnwidth\hsize\csname@twocolumnfalse\endcsname

\title{Magnetic Collapse of a Neutron Gas: No Magnetar Formation}

\author{Aurora P\'erez Mart\'{\i}nez$^{a}$, Hugo P\'erez Rojas$^{b,c}$ and Herman J. Mosquera Cuesta$^{c,d}$}
\address{$^a$ Departamento de F\'{\i}sica, Universidad Nacional de Colombia, Sede Medell\'{\i}n, Apartado A\'ereo 3804, Medell\'{\i}n, Colombia\\ $^b$ ICIMAF, Calle E No. 309, 10400 La Habana, Cuba\\$^c$ Abdus Salam International Centre for Theoretical Physics, P.O. Box 586, Strada Costiera 11, Miramare 34100, Trieste, Italy \\ $^d$ Centro Brasileiro de Pesquisas F\'{\i}sicas, Laborat\'orio de Cosmologia e F\'{\i}sica Experimental de Altas Energias \\ Rua Dr. Xavier Sigaud 150, Cep 22290-180, Urca, Rio de Janeiro, RJ, Brazil}

\date{\today}

\maketitle

\begin{abstract}
A degenerate neutron gas in equilibrium with a background of electrons
and protons in a magnetic field exerts its pressure anisotropically,
having a smaller value perpendicular than along the magnetic field.
For critical fields the magnetic pressure may produce the vanishing of the equatorial pressure of the neutron gas, and the outcome could be a transverse
collapse of the star. This fixes a limit to the fields to be observable
in stable pulsars as a function of their density. The final structure
left over after the implosion might be a mixed phase of nucleons and
meson ($\pi^{\pm,0},\kappa^{\pm,0}$) condensate (a strange star also
likely) or a black string, but no magnetar at all. 
\end{abstract}

\vskip 2pc

\narrowtext


]

\def\be{\begin{equation}}
\def\ee{\end{equation}}
Gravitational collapse occurs in a body of mass $M$ and radius $R$ when
its rest energy is of the same order of its gravitational energy, $Mc^2
\sim G M^2/R$. We would like to argue that for a macroscopic magnetized
body, e.g., composed by neutrons in an external field $B$, new physics
appears and a sort of collapse occurs, when its internal energy density
$U$ is of the same order than its magnetic energy density ${\cal M} B$,
where ${\cal M}$ is the magnetization. This problem is interesting in
the context of astrophysical and cosmological objects, as neutron
stars. Extremely magnetized neutron stars, or magnetars\cite{DT92},
have been attracting the attention of reseachers in the last
years\cite{kouveliotou98}. Their magnetic fields are
estimated to be of order of $10^{15}$
G\cite{kouveliotou98}. For fields of this order of
magnitude, there are values of the density for which the magnetic
energy of the system becomes of the same order than its total energy.
For these physical conditions, superdense matter composed of neutral
particles having a magnetic moment may undergo a transverse collapse
since its pressure perpendicular to $B$ vanishes. This implosion is
driven by the same mechanism described in \cite{Chaichian} for charged
particles. As discussed below, the resulting object may be a hybrid or
a strange star (depending on the macroscopic properties of the
imploding neutron star) but no any magnetar.
In the present paper (a preliminar report of which was presented in
\cite{Puebla}) we will approach this problem by starting from a model of a relativistic degenerate neutron gas, pervaded by an extremely strong
magnetic field, in equilibrium with a background of electrons and
protons. The last ones needed to prevent neutron beta decay $n\to p+e+\bar{\nu}$ through Pauli's principle. This configuration is maintained in approximate equilibrium through the equation $\mu _n=\mu _p+$ $\mu _e+$
$\mu_{\bar{\nu}}$ among their chemical potentials. However, as there is
a flux of neutrinos escaping from the star its chemical potential
would be taken as zero.

The Dirac equation for a charged particle having an anomalous magnetic
moment and placed in a strong magnetic field was first derived by Pauli
\cite{Pauli1940}. The free particle spectrum was obtained by Ternov
{\it et. al} \cite{Ternov}. For neutral particles, it was presented
first by Lostukov and Leventuiev, and reported by Klimenko
\cite{Klimenko}. We get for neutrons the eigenvalues,

\begin{equation} E\hspace{0in}_n(p,B,\eta )=\sqrt{p_3^2 +
(\sqrt{p^2_{\perp}+ m_n^2}+\eta qB)^2}\label{ein} 
\end{equation}

\noindent where $p_3$, $p_{\perp}$ are respectively the momentum components along and perpendicular to the external field $\bf B$, $q =1.91 M_n$, where
$M_n$ is the nuclear magneton, $\eta =1,-1$ are the $\sigma_3$
eigenvalues corresponding to the two orientations of the magnetic
moment (parallel and antiparallel) with regard to the field $B$. The
expression (\ref{ein}) shows manifestly the change of spherical to
axial symmetry with regard to  momentum components. In the
non-relativistic limit and for strong fields, this is reflected as small corrections to the spectrum as $E\hspace{0in}(p,B)=p^2/2m_n +\eta q B (1 + p_{\perp}^2/2m_n^2) + q^2 B^2/2 m_n$, valid for any neutral Fermi gas with zero electric dipole moment and nonzero magnetic moment.

The partition function ${\cal Z}$ which is obtained from the density
matrix describing the model, leads to a thermodynamic potential
$\Omega =-T\ln{\cal Z}$ involving the contributions from the species
involved, which are considered to be in chemical equilibrium among
themselves.

Having an equation relating the chemical potentials, and demanding 
conservation of both baryonic number $N_n + N_p = N_B$ and electric charge,
$N_p + N_e= 0$, one may think to solve exactly the problem in terms of
the external field as a parameter. In the present paper, however, we
focus our discussion on the properties of the equation of state.\footnote{In concluding this paper, the authors got awared of a recent paper
\cite{Mathews} on neutron gas in a magnetic field, containing
expressions similar to ours for the spectra and densities of neutrons
and protons (already reported in \cite{Puebla}), but different
equations of state. The behavior of proton-neutron fraction and pion
condensation are studied in detail in that paper.} As it is known,
the denser the Fermi gas, the better the ideal gas approximation
\cite{Landau}. This property is still valid in presence of external
fields. We thus proceed to calculate the ideal gas thermodynamical
quantities, and begin with the neutron thermodynamic potential.

For the one-loop thermodynamical potential, $\Omega_n$, we can write the 
expression 

\begin{eqnarray} 
\Omega_n &=&-\Omega_0 \sum_{\eta =1,-1}\large[\frac{x f(x,\eta y)^3} {12}+\frac{(1 + \eta y)(5\eta y - 3)x f(x,\eta y)}{24}\nonumber \\
[1em]&-&\mbox{} \frac{(1 + \eta y)^3(3 - \eta y)}{24} L
(x, \eta y)- \frac{\eta y x^3}{6}s(x, \eta y)\large], 
\end{eqnarray}

Above we have defined the variables $x = \mu_n/m_n$ and $y = q B/m_n$, and the functions 

\begin{eqnarray}
&  & f(x,\eta y)  = \sqrt{x^2 - (1+\eta y)^2},  \nonumber \\
&  & s (x, \eta y) = (\pi/2 - \sin^{-1}(1 + \eta y)/x, \nonumber \\
&  & L (x, \eta y) = \ln ( x + f(x,\eta y))/( 1+ \eta y).
\end{eqnarray}
For the density, defined as $N_n =-\partial \Omega _n/\partial
\mu _n $,  we can write

\begin{eqnarray} 
N_n &=&N_0\sum_{\eta =1,-1}\large[\frac{f(x,\eta y)^{3}} {3} + \frac{\eta y(1 + \eta y)f(x,\eta y)}{2}\nonumber \\
[1em]&-&\mbox{} \frac{\eta y x^2}{2} s (x, \eta y)\large], 
\end{eqnarray}

while for the magnetization, given by ${\cal M}_n =-\partial \Omega_n/\partial B $, we get

\begin{eqnarray} {\cal M}_n &=&-{\cal M}_0\sum_{\eta=1,-1}\eta
\large[\frac{(1- 2\eta y) xf(x,\eta y)}{6} \nonumber \\
[1em] &-&\frac{(1 + \eta y)^2 (1 - \eta y/2)}{3} L(x, \eta y) + \frac{x^3}{6} s(x, \eta y)\large], 
\end{eqnarray}
\noindent
where $N_0 =M^3_n /4\pi^2 \sim 2.04 \times 10^{39}$, $\Omega_0 =N_0 M_n
\sim 3.0 \times 10^{36}$, and ${\cal M}_0 = M_n^2 e \mu/4\pi^2 \sim 2.92 
\times 10^{16}$ and one can write ${\cal M}_n ={\cal M}_n^+ (\eta = -1)- 
{\cal M}_n^+ (\eta = +1)$, and obviously, ${\cal M}_n \geq 0$.

If we include the anomalous magnetic moment for electrons, one can give
a common formula for the spectrum of electrons and protons in the
external field $B$ as\cite{Ternov}: $E= \sqrt{p_3^2 + (\sqrt{2eBn + 
m^2_{e,p}} +\eta q_{e,p} B)^2}$, where $q_e =\alpha e/8\pi m_e$, $q_p
\sim 2.79 M_n$. For neutrons, the critical field at which the coupling 
energy of the anomalous magnetic moment equals the rest energy is 
$B_{cn} = 1.57 \times 10^{20}$. For protons $B_{cp}=2.29 \times 10^{20}$G, while for electrons it is of order $B \sim e/r_0^2 \sim 10^{-16}$G, 
being $r_0$ the classical electron radius. However, these estimates must
change if corrections depending on the field are made to the anomalous
magnetic moment. At these critical fields vacuum becomes unstable under
particle-antiparticle pair production. Such decays are, however,
suppressed in the dense gases by Pauli's principle.

By defining $x_p = \mu_p/m_p$, $y_p = q_p /m_p$, $b = 2e/m_p^2$, $q_p
= 2.79 M_n$, then $y_p = 2.79 e/2 m_p^2$. We name also 
\begin{eqnarray} 
&  & g(x_p,B,n)  =  \sqrt{x_p^2 -h(B,n)^2 }, \; {\rm and} \nonumber \\
&  & h(B,n)=(\sqrt{b B n + 1}  + \eta y_p B),
\end{eqnarray}

\noindent to have for the star's proton thermodynamical potential
\begin{eqnarray} 
\Omega_ p & = & - \frac{eB m_p^2}{4 \pi^2}\sum_n
\sum_{\pm \eta}[x_p g(x_p, B,n)- h(B,n)^2 \nonumber \\[1em] &\times &
\ln \frac{x_p+g(x_p,B,n)}{h(B,n)}, 
\end{eqnarray}

and for its density
\begin{equation} 
N_p= \frac{eB m_p}{2 \pi^2}\sum_n \sum_{\pm \eta}
g(x_p, B, n).  
\end{equation}

The magnetization is given by
\begin{eqnarray} 
{\cal M}_p & = & \frac{e m_p^2}{4 \pi^2}\sum_n
\sum_{\pm \eta}\large[x_p g(x_p, B,n) - [ h(B,n)^2 + (\eta y_p
\nonumber \\
[1em] & + & (bn/2\sqrt{bBn + 1}))]\ln \frac{ x_p +\sqrt{x_p^2- g(x_p, B,n)}}{h(B,n)}, 
\end{eqnarray}

\noindent where the coefficents of these formulae are $N_0 =e m_p B /2\pi^2 \sim 4.06 \times 10^{19}B$, $\Omega_0 =N_0 m_p B \sim 6.1 \times 10^{16}B$,
and ${\cal M}_0 = N_0 m_p =\Omega_0/B$.

For low magnetic fields $n_{max}$ is very large, and one can
approximate the sum over Landau quantum numbers by an integral. This
would lead to a neutron to proton ratio $ N_n/N_p$ $\geq 8,$ similarly
as the zero magnetic field case \cite{Weinberg}. The maximum occupied
Landau quantum number $n$ may be given as $n_{max}= (x_p - \eta y_p B)^2 -1/bB$. 

For $B \ll B_{cp}$, so that $y_p B \ll 1$, and $x_p \geq 1$, one can take
approximately $n_{max} \sim (x_p^2 -1)/bB$, and for fields large enough
$n_{max} = 0$. As $x_p \sim x_n$, the proton density decreases with
increasing $B$, favoring the inverse beta decay. In the extreme case of
confinement to the Landau ground state we can estimate the bound $N_p
=N_e \leq 2.5 \times 10^9 B N_n^{1/3}$. For this limit it is roughly
$\Omega_n/\Omega_p \geq 4 \times 10^{-10}N^{2/3}/B$. For fields $B
\sim m_p/q_p$ and $x_p \gg 1$, $n_{max}\geq 1$, and thus large Landau
numbers are again occupied. However, the dominant longitudinal pressure
and magnetization comes from the neutron gas.

The lost of rotational symmetry of the particle spectrum determines an
anisotropy in the thermodynamic properties manifesting in different
equations of state for directions parallel and perpendicular to the
external field, as is seen from the energy-momentum tensor in the
constant magnetic field ${\cal T}_{\mu\nu }$ \cite{Hugop}, \cite{Chaichian}. Its spatial components ${\cal T}_{11}={\cal T}_{22}=P_{\perp}$ and ${\cal T}_{33}= P_3$, which
contain the sum of the partial pressures  of the several 
species involved, are
\begin{equation}  
P_{\perp }=-\Omega _{}-B{\cal M}\hspace{1cm}P_3=-\Omega, \label{balance}
\end{equation}
\noindent where $\Omega = \sum_i\Omega_i$ is the total
thermodynamical potential, $i=n,p,e$, and ${\cal M=}\sum_i{\cal M}_i$. By calling $S = \sum_i S_i$ and $N = \sum_i N_i$, where both the partial entropy and density 
are $S_i =\partial \Omega _i/\partial T$, $N_i=\partial \Omega
_i/\partial \mu _i$, one can write the internal energy density as
$U= -{\cal T}_{44}= \mu N + TS + \Omega$ (we take $T=0$ in the present
degenerate case), where the total quantities $U$, $\mu N$, $\Omega$ are
of similar order. For positive magnetization, which is our case, the
transverse pressure exerted by the magnetized particles is smaller than
the longitudinal one in the amount $B{\cal M}$. 
If we assume that the body is in equilibrium
under the balance of neutron and gravitational pressures, being the
latter of order $P_{grav} \sim GM^2/R^4$, where $R$ is the geometric
average radius of the star, the body stretch along the direction of the
magnetic field. (This effect can be figured out from looking at the spectrum described by Eq.(\ref{ein}), since the contribution from $\eta =-1$ terms are dominant. Then, if one approximates the second term inside the square root as $[m_n + (p^2_{\perp}/2m_n -qB)]^2$, the term in parenthesis accounts for the transverse kinetic energy. This term decreases as the magnetic field increases. Equating it to zero, by taking $p_{\perp} \sim p_F$, where $p_F$ is the Fermi momentum, we obtain a functional relation between $\mu_n$ and $B$ leading to the vanishing of the transverse kinetic energy, and in consequence of the transverse pressure. Observe that for $x=1.001$, $y \sim 10^{-6}$. A more accurate result is obtained, however, from ${\cal T}_{\perp}=0$). 
This extreme case of vanishing of $P_{\perp}=-\Omega -B{\cal
M}$  means that the transverse gravitational and Fermi gas
pressures cannot compensate each other and an instability appears leading to a transverse collapse. We do not enter in the quantitative study of this
problem, which would lead to a special sort of hybrid stars\cite{glendenning,madsen99} or black strings
\cite{Chiba,Hayward}. In Figure 1 we have drawn the
equation $P_{\perp n}=0$ in terms of the variables $x,y$. We observe that 
there is a continuous range of values of the
chemical potential, starting from values $x=1.001$  and magnetic field
intensities, from $y=0.000001$, for which the collapse
takes place. The latter value of $y$ means fields on 
order of $B \sim 10^{14}$ G. To these ranges of $x,y$ corresponds a
continuous range of densities, from $10^{-4}N_0$ ($10^{11}$
g/cm$^{-3}$) onwards. Thus, although stable neutron stars start to occur for densities $N \sim 10^{-2}N_0$\cite{shapiro}, the mechanism of decreasing the
transverse pressure may act in the external regions of them, where densities of order $10^{-4}N_0$ may occur. This would lead
to transverse compression of the whole mass of the star, concentrating the magnetic lines of force and increasing $B$. For $B \sim 
B_{cp,cn}$, the magnetic coupling of quarks with $B$
becomes of the order of their binding energy through the colour field.
This might lead to a deconfinement phase transition leading to a quark
(q)-star, a pressure-induced transition to uds-quark matter via ud-quark 
condensates, as discussed in Refs.(\cite{madsen99,glendenning}). 

\begin{figure} 
\centering{\epsfig{file=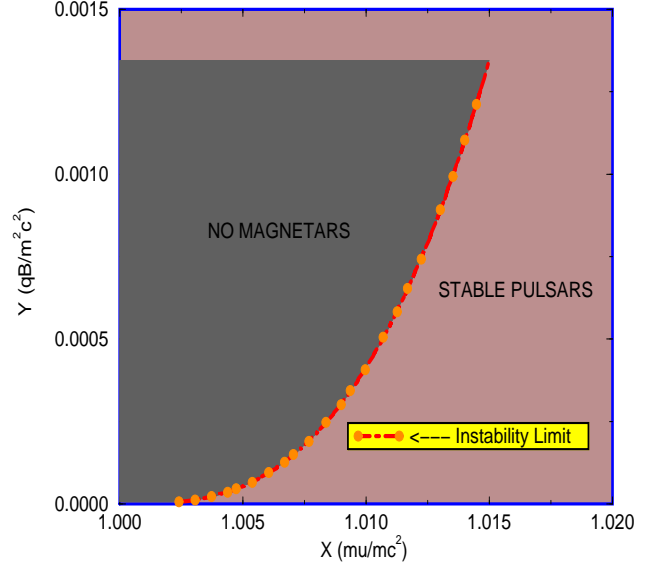,height=8.8cm,width=8.8cm}}
\caption{The instability condition: $P_{\perp} = 0$. A neutron star having a configuration such that its dynamical stage would be represented by a point above the central curve in this plot would be unstable to transversal collapse,  since $P_{\perp} \leq 0$ there.}
\end{figure}

Observations of GRB980827 from SGR 1900+14 evidenced the existence of 
a stable pulsation with period 5.16 s\cite{hurley99-1}. In analogy 
to the case for SGR 1806-20\cite{kouveliotou98}, the
observed spin-down rate of the pulse period, $\dot{P} = 1.1 \times
10^{-10}$ s s$^{-1}$, led Kouveliotou et al.\cite{kouveliotou98} to
announcing the discovery of a {\it magnetar} in the source SGR 1900+14. 
These findings apparently lead to verify the Duncan and Thompson\cite{DT92}
magnetar model for SGRs. Kouveliotou et al.\cite{kouveliotou98}
claimed that spin down may be explained by the emission of dipolar
radiation from a NS endowed with a very strong magnetic field $B \sim
[2-8] 10^{14}$ G.  Next we briefly review the standard theory of
magnetars. We shall show why they cannot survive after reaching the
claimed superstrong magnetic fields, and then we present prospectives
for a hybrid or strange star to appear as a remnant of the quantum
collapse.
According to Duncan and Thompson\cite{DT92}, neutron stars (NSs) with
very high dipole magnetic field strength, $B_D \sim [10^{14}-10^{15}]$
G, may form when (classical) conditions for a helical dynamo action are
efficiently met during the seconds following the core-collapse in a
supernova (SN) explosion\cite{DT92}. A newly-born NS may undergo
vigorous convection during the first 30 s following its formation. If
the NS spins (differentially) sufficiently fast ($P \sim 1$ms) the
conditions are created for the $\alpha-\Omega$ dynamo action to be
built. Collapse theory shows that some presupernova stellar cores could
endow enough spin so as to rotate near their Keplerian equatorial
velocity (the break-up spin) $\Omega_K \geq ([\frac{2}{3}]^3 GM/R^3)^{1/2}$
after core bounce. Under these conditions, fields as large as $B\sim
10^{17} (\frac{P}{1 {\rm ms}}$) G may be generated as long as the
differential rotation is dragged out by the growing magnetic
stresses\cite{DT92}.  For this process to efficiently operate the ratio
between the spin rate ($P$) and the convection overturn time scale
($\tau_{con}$), the Rossby number $R_0$, should be $\leq 1$ ($R_0 \gg
1$ should induce less effective mean-dynamos\cite{DT92}). In this
case,  amplification of the magnetic field strength by these helical
motions is not precluded, since the $\alpha^2$ or $\alpha-\Omega$
dynamos may survive depletion due to turbulent diffusion. An ordinary
dipole $B_{sat} \sim [10^{12}-10^{13}]$ G may be built by incoherent
superposition of small dipoles of characteristic size $\lambda \sim
[\frac{1}{3} - 1]$km and $B_{sat} = (4\pi \rho)^{1/2} \lambda
/\tau_{con} \simeq 10^{16}$ G. At such fields, the huge rotational energy
of a $f \geq 1$kHz NS is leaked out  via {\it magnetic braking}, and
an enormous energy is injected into the SN remnant. This energy may power  
a {\it plerion} in the SN remnant.  

As shown above, at the end we are left with a NS with an extremely high
field strength and a huge matter density $\rho \sim [10^{14}-10^{12}]$
gcm$^{-3}$. As illustrated in Figure 1, those are the conditions for
the quantum instability to start to dominate the dynamics of the young
pulsar. At this stage, the magnetic pressure inwards may overpass the star's energy density at its equator, as defined in Eq.(\ref{balance}), and the collapse becomes unavoidable. As the collapse proceeds, higher and
higher densities are reached till the point the supranuclear density may 
reverse the direction of implosion (hybrid star). From that moment, the 
sound wave generated at the core bounce builds itself into a shock wave 
traveling through the star at $V_{SW} = c/\sqrt{3}$ kms$^{-1}$. Although the magnetic field strength could be quite large as the collapse advances,
the kinetic energy ($E\sim 10^{51}$ erg, the mean energy obtained in
simulations of SN driven by the {\it prompt shock}\cite{mueller99}) carried away by the shock wave may counterbalance it, and even surpase
it, i.e., its {\it ram} pressure will equal the magnetic pressure at   
$r_A$, the Alfv\'en radius given by
\be 
\rho_{eject} V_{SW}^2 \geq \frac{B^2}{8\pi\mu_0} \left(\frac{R}{r_A}\right)^6,\; r_A = \left(\frac{2\pi^2}{G\mu_0^2}\right)^{1/7} \left[\frac{B^4 R^{12}}{M\dot{m}^2}\right]^{1/7}, 
\ee

Then the magnetic field lines are pushed out, and finally broken, from 
$r_A$ onwards, into the SN remnant surroundings as a violent explosion that 
dissipates a large part of the magnetic flux $(\Phi \sim B^2r_A^2)$ 
trapped inside the magnetar
magnetosphere.  This is analogous to the mechanism operating during a
solar flare or a coronal mass-ejection, where the very high $B$ in the
"sun-spot" is drastically diminished.  Although the process is quite
fast, the large amount of matter ejected from the star at such large
velocities drains out the dipole field of the remnant below the quantum
electrodynamic limit of $B_{QED} \sim 10^{13}$G. Further, since all the
differential rotation has been dragged up to build up the superstrong
magnetic field, then nothing else remains to make it to grow to its
precollapse value.  Thus no such ultra high $B$ should reappear. We 
may be left with a submillisecond strange star\cite{madsen99} or a
hybrid star\cite{glendenning} with "canonical" field strength.

We conclude that if a degenerate neutron gas is under the action of a
strong magnetic field $B\lesssim B_c$, for adequate values of its
density its transverse pressure vanishes, the outcome being a
transverse collapse. This phenomenon establishes a limit to the
magnetic field expected to be observable in a neutron star, as a
function of its density, and suggests a possible end in the evolution
of the highly magnetized neutron star as a  mixed phase of nucleons and
$\pi^{\pm,0},\kappa^{\pm,0}$ meson condensate or a black string. Thus this 
result implies that most of the observed pulsars should have some either 
strange matter or meson condensate in their interiors.

\end{document}